# A Note on the Use of the Woodbury Formula To Solve Cyclic Block Tri-Diagonal and Cyclic Block Penta-diagonal Linear Systems of Equations

Milan Batista

University of Ljubljana, Faculty of Maritime Studies and Transport

Pot pomorščakov 4, 6320 Portorož, Slovenia, EU

milan.batista@fpp.edu

Abdel Rahman A. Ibrahim Karawia

Mathematics Department, Faculty of Science, Al-Qassim University,

Buraidah 51452, Al-Qassim, KSA.

abibka@mans.edu.eg

**Abstract**

The article presents the theoretical background of the algorithms for solving cyclic block tridiagonal and cyclic block penta-diagonal systems of linear algebraic equations present in [1] and [2]. The theory is based on the Woodbury formula.

**1 Introduction**

The algorithms for solving the cyclic block tri-diagonal (CBTS) and cyclic block penta-diagonal systems (CBPS) of equations presented in ref [1], [2] are based on ad hoc introduction of the new unknown vector(s) which transform the original system into a non-cyclic system which can then be solved by standard methods. In this short paper the theoretical background of the algorithms based on the Woodbury formula ([3],[4]) will be given. Before proceeding it is perhaps necessary to emphasize that while the present treatment for solving CBTS differs only slightly from the method present in [5], the method for solving CBPS seems to be different from known methods.

Now, the object of consideration is the cyclic block penta-diagonal system of linear algebraic equations

$$Ax = f \tag{1}$$

where

$$A = \begin{bmatrix} B_1 & C_1 & D_1 & 0 & \cdots & E_1 & A_1 \\ A_2 & B_2 & C_2 & D_2 & \cdots & 0 & E_2 \\ E_3 & A_3 & B_3 & C_3 & \cdots & \vdots & \vdots \\ \vdots & \vdots & \ddots & \ddots & \ddots & \ddots & \vdots \\ D_{n-1} & 0 & \cdots & \cdots & A_{n-1} & B_{n-1} & C_{n-1} \\ C_n & D_n & \cdots & \cdots & E_n & A_n & B_n \end{bmatrix} \tag{2}$$

is a cyclic block penta-diagonal system matrix, $x = (x_1, x_2, ..., x_n)^{\mathrm{T}}$ and $f = (f_1, f_2, ..., f_n)^{\mathrm{T}}$ are unknown and known vectors (RHS vector), respectively, and $n \geq 4$ is the number of equations. $A_k$, $B_k$, $C_k$, $D_k$ and $E_k$ are matrices of size $m \times m$, and $f_k$ and $x_k$ are vectors of size $m$. In what follows, if not stated otherwise, all other matrices have size $m \times m$ and all other vectors have size $m$. Also the vectors and matrices will be denoted by Roman lower-case and upper case letters, and scalars will be denoted by Greek lower-case letters. A unit matrix will be denoted as $I$ and its dimension should be clear from the context; also from the context the meaning of the sign 0, which will be used for denoting scalar value or a null-matrix, should be clear.

**2 Cyclic block tri-diagonal system**

When $D_k = E_k = 0$ the system (1) becomes a CBTS. To solve it the system matrix $A$ is first de-composed into the form suggested by the Woodbury formula

$$A = \tilde{A} + UV^T \tag{3}$$

where is $\tilde{A}$ is an $n \times n$ block matrix, and $U$ and $V$ are $n \times 1$ block matrices. For systems under discussion a simple way to restore the off diagonal terms in (2) by (3) is determining that the product of the matrices $U$ and $V^T$ in (3) is of the form

$$UV^T = \begin{bmatrix} U_1 \\ 0 \\ \vdots \\ 0 \\ U_n \end{bmatrix} \begin{bmatrix} V_1 & 0 & \cdots & 0 & V_n \end{bmatrix} = \begin{bmatrix} X_1 & 0 & \cdots & 0 & A_1 \\ 0 & 0 & \cdots & 0 & 0 \\ \vdots & \vdots & \cdots & \vdots & \vdots \\ 0 & 0 & \cdots & 0 & 0 \\ C_n & 0 & \cdots & 0 & X_n \end{bmatrix} \tag{4}$$

where matrices $U_1$, $U_n$, $V_1$, $V_n$, $X_1$, and $X_n$ should satisfy the following conditions

$$U_1 V_n = A_1 \qquad U_n V_1 = C_n \qquad U_1 V_1 = X_1 \qquad U_n V_n = X_n \tag{5}$$

By introducing (4) into (3) one finds

$$\tilde{A} = \begin{bmatrix} B_1 - X_1 & C_1 & 0 & 0 & \cdots & 0 & 0 \\ A_2 & B_2 & C_2 & 0 & \cdots & 0 & 0 \\ 0 & A_3 & B_3 & C_3 & \cdots & \vdots & \vdots \\ \vdots & \vdots & \ddots & \ddots & \ddots & \ddots & \vdots \\ 0 & 0 & \cdots & \cdots & A_{n-1} & B_{n-1} & C_{n-1} \\ 0 & 0 & \cdots & \cdots & 0 & A_n & B_n - X_n \end{bmatrix} \tag{6}$$

The system (5) is four equations for six unknown matrices so its solution is not unique. A simple way to satisfy the first two of (5) is to choose

$$U_1 = I/\alpha \qquad V_n = \alpha A_1 \qquad U_n = I/\gamma \qquad V_1 = \gamma C_n \tag{7}$$

where $\alpha$ and $\gamma$ are non-zero scalars. Consequently,

$$X_1 = \frac{\gamma}{\alpha} C_n \qquad X_n = \frac{\alpha}{\gamma} A_1 \qquad (8)$$

Here the present derivation differs from that in [5] where the values $\alpha = \gamma = 1$ are implicitly assumed. By (8) the matrix $\tilde{A}$ takes the same form as the matrix of the system (3) in ref [1]. By using (3) the system (1) becomes

$$\left(\tilde{A} + UV^T\right)x = f$$

Multiplying this with $\tilde{A}^{-1}$, where $\tilde{A}$ is assumed to be non-singular, yields the system

$$\left(I + ZV^T\right)x = y \qquad (9)$$

where $Z = [Z_1,...,Z_n]^T$, with matrices $Z_k$ and $y = (y_1, y_2,...,y_n)^T$, with vectors $y_k$, are given by

$$Z = \tilde{A}^{-1} U \qquad y = \tilde{A}^{-1} f \qquad (10)$$

This is precisely the solution of the systems (6) and (7) used in ref [1]. From (9) the final solution of (1) is

$$x = \left(I + ZV^T\right)^{-1} y \qquad (11)$$

where $I + ZV^T$ is assumed to be non-singular. With the Woodbury formula the inversion of this matrix is

$$\left(I + ZV^T\right)^{-1} = I - ZM^{-1}V^T \qquad (12)$$

where

$$M = I + V^T Z \tag{13}$$

is a matrix of order $m \times m$ which is assumed to be nonsingular. By (11) and (12) the solution $x$ can be written in the form (eq 5 in [1])

$$x = y - Zu \tag{14}$$

where $u = M^{-1} V^T y$ is the vector. By using (7) one can easily find that $M = I + \gamma C_n Z_1 + \alpha A_1 Z_n$ and $V^T y = \gamma C_n y_1 + \alpha A_1 y_n$ so the vector $u$ is computed by

$$u = \left(I + \gamma C_n Z_1 + \alpha A_1 Z_n\right)^{-1} \left(\gamma C_n y_1 + \alpha A_1 y_n\right) \tag{15}$$

This is precisely the same vector as vector (15) in ref [1]. In this way the equivalence between the method used in [1] and when using the Woodbury formula is demonstrated.

**3 Cyclic block penta-diagonal system**

When $E_1 = E_2 = D_{n-1} = D_n$ the CBPS (1) can be solved by the same method as the CBTS. In general, however, to solve a CBPS the system matrix $A$ may be de-composed into the following form

$$A = \tilde{A} + UV^T + PQ^T \tag{16}$$

where $\tilde{A}$ is an $n \times n$ block matrix, and $U$, $V$, $P$ and $Q$ are $n \times 1$ block matrices. To restore the off diagonal terms in (2) the product of the matrices $U$ and $V^T$ should be

$$UV^T = \begin{bmatrix} U_1 \\ 0 \\ \vdots \\ 0 \\ U_n \end{bmatrix} \begin{bmatrix} V_1 & V_2 & \cdots & V_{n-1} & V_n \end{bmatrix} = \begin{bmatrix} X_1 & Y_1 & \cdots & E_1 & A_1 \\ 0 & 0 & \cdots & 0 & 0 \\ \vdots & \vdots & \cdots & \vdots & \vdots \\ 0 & 0 & \cdots & 0 & 0 \\ C_n & D_n & \cdots & Y_n & X_n \end{bmatrix} \quad (17)$$

where matrices $U_1$, $U_n$, $V_1$, $V_2$, $V_{n-1}$, $V_n$, $X_1$, $Y_1$, $X_n$, and $Y_n$ should satisfy the following conditions

$$U_1 V_n = A_1 \quad U_n V_1 = C_n \quad U_1 V_{n-1} = E_1 \quad U_n V_2 = D_n$$
$$U_1 V_1 = X_1 \quad U_1 V_2 = Y_1 \quad U_n V_{n-1} = Y_n \quad U_n V_n = X_n \quad (18)$$

and the product of matrices $P$ and $Q^T$ should yield

$$PQ^T = \begin{bmatrix} 0 \\ P_2 \\ \vdots \\ P_{n-1} \\ 0 \end{bmatrix} \begin{bmatrix} Q_1 & 0 & \cdots & 0 & Q_n \end{bmatrix} = \begin{bmatrix} 0 & 0 & \cdots & 0 & 0 \\ X_2 & 0 & \cdots & 0 & E_2 \\ \vdots & \vdots & \cdots & \vdots & \vdots \\ D_{n-1} & 0 & \cdots & 0 & X_{n-1} \\ 0 & 0 & \cdots & 0 & 0 \end{bmatrix} \quad (19)$$

where matrices $P_2$, $P_{n-1}$, $Q_1$, $Q_n$, $X_2$ and $X_{n-1}$ should satisfy the conditions

$$P_2 Q_n = E_2 \quad P_{n-1} Q_1 = D_{n-1} \quad P_2 Q_1 = X_2 \quad P_{n-1} Q_n = X_{n-1} \quad (20)$$

By knowing $U$, $V$, $P$ and $Q$, one, from (16), finds that $\tilde{A} = A - UV^T - PQ^T$ or explicitly

$$\tilde{A} = \begin{bmatrix} B_1 - X_1 & C_1 - Y_1 & D_1 & 0 & \cdots & 0 & 0 \\ A_2 - X_2 & B_2 & C_2 & D_2 & \cdots & 0 & 0 \\ E_3 & A_3 & B_3 & C_3 & \cdots & \vdots & \vdots \\ \vdots & \vdots & \ddots & \ddots & \ddots & \ddots & \vdots \\ 0 & 0 & \cdots & \cdots & A_{n-1} & B_{n-1} & C_{n-1} - X_{n-1} \\ 0 & 0 & \cdots & \cdots & E_n & A_n - Y_n & B_n - X_n \end{bmatrix} \quad (21)$$

Both systems (18) and (20) are under-determinate. A simple way to satisfy the first two of (18) is to choose

$$U_1 = I/\alpha \qquad V_n = \alpha A_1 \qquad U_n = I/\gamma \qquad V_1 = \gamma C_n \qquad (22)$$

where $\alpha$ and $\gamma$ are a non-zero scalars. From the rest of (18) one obtains

$$V_{n-1} = \alpha E_1 \qquad V_2 = \gamma D_n$$

$$X_1 = \frac{\gamma}{\alpha} C_n \qquad Y_1 = \frac{\gamma}{\alpha} D_n \qquad Y_n = \frac{\alpha}{\gamma} E_1 \qquad X_n = \frac{\alpha}{\gamma} E_1 \qquad (23)$$

Similarly, to satisfy first two of (20) one may set

$$P_2 = I/\beta \qquad Q_n = \beta E_2 \qquad P_{n-1} = I/\delta \qquad Q_1 = \delta D_{n-1} \qquad (24)$$

where $\beta$ and $\delta$ are non-zero scalars, and, consequently,

$$X_2 = \frac{\delta}{\beta} D_{n-1} \qquad X_{n-1} = \frac{\beta}{\delta} E_2 \qquad (25)$$

By (23) and (25) the matrix (21) has the form of matrix (5) in [2]. Now, by (16) the system (1) becomes

$$\left(\tilde{A} + UV^T + PQ^T\right)x = f \qquad (26)$$

By multiplying this by $\tilde{A}^{-1}$, where $\tilde{A}$ is assumed to be non-singular, one obtains the system

$$\left(I + ZV^T + WQ^T\right)x = y \qquad (27)$$

where $Z = [Z_1,...,Z_n]^T$, with matrices $Z_k$, $W = [W_1,...,W_n]^T$, with matrices $W_k$ and $y = (y_1, y_2,..., y_n)^T$, with vectors $y_k$, are given by

$$Z = \tilde{A}^{-1}U \qquad W = \tilde{A}^{-1}P \qquad y = \tilde{A}^{-1}f \tag{28}$$

This is again precisely the solution of the systems (9) in ref [2]. From (27) the final solution of (1) is

$$x = \left(I + ZV^T + WQ^T\right)^{-1} y \tag{29}$$

where $I + ZV^T + WQ^T$ is assumed to be non-singular. By nothing that $ZV^T + WQ^T = [Z,W]\begin{bmatrix}V^T, Q^T\end{bmatrix}^T$ the inversion of this matrix can be, by use of Woodbury formula, written as (this was corrected)

$$\left(I + ZV^T + WQ^T\right)^{-1} = I - [Z,W]M^{-1}\begin{bmatrix}V^T, Q^T\end{bmatrix}^T \tag{30}$$

where

$$M = \begin{bmatrix} I + V^T Z & V^T W \\ Q^T Z & I + Q^T W \end{bmatrix} \tag{31}$$

is a matrix of the order $2m \times 2m$ which is assumed to be nonsingular. The solution $x$ can therefore be written in the form (eq 8 in [2])

$$x = y - Zu - Wv \tag{32}$$

where $u = M^{-1}V^T y$ and $v = M^{-1}Q^T y$ are the vectors. By using (22) and (24) and (31) one can easily restore the system (17) in [2] for the determination of $u$ and $v$. The equivalence between the method used in [2] and that using the generalized Woodbury formula is thus shown.

**4 Conclusions**

In the article the theoretical background of the method for solving cyclic block tri- and penta-diagonal linear systems present in ref [1] and [2] are given. In these references future details, implementation and numerical examples can be found. As is well known ([4],[5]) the attractiveness of the present method is that the cyclic block system (or any other system with the same off diagonal structure) can be solved through fewer operations than by using direct elimination or some direct sparse system algorithms (for example sparse 'skyline') if one poses an efficient way to solve the corresponding block diagonal system, for, besides solving this system, one must for CBTS additionally solve just the system of order $m \times m$ and in the case of CBPS just the system of order $2m \times 2m$. Clearly, the present method works if the matrices $\tilde{A}$ -- (12) and (31) -- and $M$ --(13) and (31)-- are non-singular. From the practical point of view non-singularity of matrices is tested by the program. Theoretical consideration is beyond the scope of the paper. The same is true also for the discussion of the optimal choice of parameters $\alpha, \beta, \gamma$ and $\delta$, which remains an open question.